\documentclass[twocolumn]{aastex631}

\begin{document}

\title{Using SOFIA's EXES to search for C$_6$H$_2$ and C$_4$N$_2$ in Titan's atmosphere}

\author{Zachary C. McQueen}
\affiliation{NASA Goddard Space Flight Center, 8800 Greenbelt Rd,
Greenbelt, MD 20771, USA}
\affiliation{Southeast Universities Research Association, 
1201 New York Avenue, Suite 430, Washington, DC 20005, USA}

\author{Conor A. Nixon}
\affiliation{NASA Goddard Space Flight Center, 8800 Greenbelt Rd,
Greenbelt, MD 20771, USA}

\author{Curtis de Witt}
\affiliation{Space Science Institute,
4765 Walnut St STE B, 
Boulder, CO 80301,USA}

\author{Véronique Vuitton}
\affiliation{Univ. Grenoble Alpes, CNRS, IPAG, 38000 Grenoble, France}

\author{Panayotis Lavvas}
\affiliation{GSMA UMR 7331, University Reims Champagne Ardenne, Reims, France}

\author{Juan Alday}
\affiliation{Instituto de Astrofísica de Andalucía, CSIC, Granada, Spain}

\author{Nicholas A. Teanby}
\affiliation{University of Bristol, Queens Road Clifton, Bristol, BS8 1RJ, United Kingdom}

\author{Joseph Penn}
\affiliation{Oxford University, Clarendon Laboratory, Parks Rd, Oxford OX1 3PU, United Kingdom}

\author{Antoine Jolly}
\affiliation{Université Paris Est Créteil and Université Paris Cité, Créteil, France}

\author{Patrick G. J. Irwin}
\affiliation{Oxford University, Clarendon Laboratory, Parks Rd, Oxford OX1 3PU, United Kingdom}



\begin{abstract}

In Titan's atmosphere, the chemistry of small hydrocarbons and nitriles represent an important link from molecular species to the ubiquitous organic haze that gives Titan its characteristic yellow color. Here we present a new search for two previously undetected molecules, triacetylene (C$_{6}$H$_{2}$) and the gas phase dicyanoacetylene (C$_{4}$N$_{2}$), using the Echelon-Cross-Echelle Spectrograph (EXES) instrument aboard the SOFIA (Stratospheric Observatory For Infrared Astronomy) aircraft. We do not detect these two molecules but determine upper limits for their mixing ratios and column abundances. We find the $3\sigma$ upper limits on the uniform volume mixing ratio (VMR) above 100 km for C$_{6}$H$_{2}$ to be $4.3\times10^{-11}$ which is lower than the photochemical model predictions. This new upper limit suggests that the growth of linear molecules is inhibited. We also put a strict upper limit on the uniform VMR for gas phase C$_{4}$N$_{2}$ above 125 km to be $1.0\times10^{-10}$. This upper limit is well below the saturation mixing ratio at this altitude for C$_{4}$N$_{2}$ and greatly limits the feasibility of C$_{4}$N$_{2}$ forming ice from condensation.

\end{abstract}



\section{Introduction}\label{sec:intro}

\begin{deluxetable*}{ccccccc}[ht!]
    \tablecaption{Description of EXES observations of Titan's atmosphere.}
    \label{tab:obs}
    \tablehead{
        \colhead{Target, setting (cm$^{-1}$)} & \colhead{Date (yyyy-mm-dd)} & \colhead{Altitude (ft)} & \colhead{Airmass} &  \colhead{Distance (A.U.)} & \colhead{$D_v$ ($km/s$)} & \colhead{Integration time ($s$)}   
    }
    \startdata
    Callisto, 621 & 2021-06-09 & 43000 & 1.84 & -      &  -       & 1408 \\ 
    Titan, 621    & 2021-06-09 & 43000 & 1.68 & 9.321  &  -20.994 & 1504 \\ 
    Vesta, 472    & 2021-06-11 & 39000 & 1.68 & -      &  -       & 1504 \\ 
    Titan, 472    & 2021-06-15 & 43000 & 1.84 & 9.264  &  -18.930 & 1600 \\
    Titan, 472    & 2021-06-17 & 43000 & 1.67 & 9.240  &  -22.499 & 2560 \\
    Titan, 472    & 2021-12-04 & 36000 & 2.00 & 10.424 &  +28.940  & 3392 \\
    \enddata
\end{deluxetable*}

Titan's major atmospheric components, N$_2$ and CH$_4$, photo-dissociate and ionize in Titan's ionosphere, consequentially leading to the formation of other hydrocarbon and nitrile species \citep{Horst.2017, Nixon.2024}. These minor components continue this photochemical process, leading to larger and more complex organic species and ultimately leading to the formation of Titan's characteristic haze \citep{Trainer.2006,Waite.2007}. Discovery of Titan's gaseous atmosphere came from early observations of Titan from Earth based astronomy, where \cite{Kuiper.1944} discovered the presence of nitrogen. Then, the Voyager flyby provided the closest infrared observations of Titan's atmosphere, allowing discovery of a wealth of new molecules \citep{Hanel.1981,Maguire.1981,Samuelson.1983,Letourneur.1993}. Space based observatories, like the Infrared Space Observatory, also contributed greatly to discovering new trace gases in Titan's atmosphere including water and benzene \citep{Coustenis.1998, Coustenis.2003,Encrenaz.2003}. The return to Titan with the Cassini mission provided excellent spatial and temporal coverage of Titan's atmosphere \citep{Coustenis.2010,Nixon.2019}, allowing for detailed IR studies of Titan over several seasons \citep{Teanby.2008a,Teanby.2019}. Following the Cassini mission, Earth-based observations at infrared and sub-millimeter wavelengths have been a major contributor to searching for and detecting trace gases that are predicted to be present in Titan's atmosphere at low abundances \citep{Palmer.2017,Thelen.2019,Lombardo.2019,Nixon.2020}. Most recently, observations using the James Webb Space Telescope have allowed for the recent discovery of the CH$_3$ radical in Titan's atmosphere \citep{Nixon.2025}.

Polyynes or poly-acetylenes are linear chains of acetylene groups and bear the molecular formula C$_{2n}$H$_2$ where $n \geq 2$. Polyynes represent one potential pathway to form macromolecules in Titan's atmosphere  that bridge the gap between simple molecules and Titan's aerosols \citep{Chassefiere.1995, Lebonnois.2002}. Polyynes form readily through the addition of the ethynyl radical (C$_2$H) to the precursor acetylene (C$_2$H$_2$) or poly-acetylene molecule i.e. diacetylene C$_4$H$_2$ \citep{Gu.2009}. Although polyynes have been observed in the interstellar medium \citep{Cernicharo.2001a, Cernicharo.2001b}, and are predicted to form in Titan's atmosphere at observable abundances \citep{Gu.2009, Vuitton.2019}, C$_4$H$_2$ is the highest order polyyne that has been detected in Titan's atmosphere \citep{Kunde.1981, Teanby.2009}. Triacetylene (C$_{6}$H$_{2}$), which forms from C$_4$H$_2$, is predicted in photochemical models to have a stratospheric abundance of $0.6$ ppb \citep{Vuitton.2019} yet, to this date the spectroscopic detection of C$_6$H$_2$ has eluded the community. 

There have been two previous studies of Titan's C$_{6}$H$_{2}$ upper limits. The first upper limit of C$_{6}$H$_{2}$, reported by \cite{Delpech.1994}, was determined by comparing the expected signal intensity from the strongest vibrational mode of C$_{6}$H$_{2}$ to the observed emission features of a reference compound and was found to be $0.6$ ppb above 100 km. This study was in agreement with the photochemical model predictions at the time, but was limited by the low spectral resolution of Voyager's IRIS and the authors note that a higher resolution spectrometer may allow C$_{6}$H$_{2}$ to be detected. \cite{Shindo.2003} updated measurements of the C$_{6}$H$_{2}$ IR spectrum and re-calculated the upper limits and lowered them to $0.44$ ppb above 150 km using a similar method. Still, at these upper limits, C$_{6}$H$_{2}$ and higher order polyynes represent a potential pathway to the formation of large hydrocarbons in Titan's atmosphere. Therefore, an updated analysis of the upper limits for C$_{6}$H$_{2}$ at high spectral resolution will provide useful estimations of the potential for higher-order polyyne formation in Titan's atmosphere.

Dicyanoacetylene (C$_4$N$_2$) in the gas phase is predicted to form in Titan's atmosphere through the reaction of HC$_3$N with CN \citep{Vuitton.2019}. This was recently shown to occur in crossed molecular beam experiments by \cite{Aragao.2025}. Additionally, C$_4$N$_2$ can form through HCN reaction with C$_3$N and through hydrogen elimination of HC$_4$N$_2$, though HC$_4$N$_2$ is formed following C$_4$N$_2$ formation \citep{Loison.2015}. C$_4$N$_2$ is unique in Titan's atmosphere because, although the gas phase species has not been detected, C$_4$N$_2$ ice has been reported in both Voyager's Infrared Interferometer Spectrometer and Radiometer (IRIS) spectra and Cassini's Composite Infrared Spectrometer (CIRS) spectra \citep{Samuelson.1997, Anderson.2016}. Following the Voyager flyby, an emission feature at  478 cm$^{-1}$ was assigned to the ice-phase $\nu_8$ vibrational mode of C$_4$N$_2$, however, the gas phase emission at 471 cm$^{-1}$ was not present. Some years later, the same anomalous feature at 478 cm$^{-1}$ arose in CIRS observations of Titan's atmosphere and was also assigned to C$_4$N$_2$ ice. Since then, there have been several studies attempting to understand why this ice feature is so strong but the gas is not present in detectable amounts.  \cite{Samuelson.1997} suggested that the gas phase C$_4$N$_2$ is enriched during the winter, leading to condensation, and then photochemically consumed rapidly at the Spring Equinox while the condensate progressively builds up in the lower stratosphere. This however was suggested to not be feasible by \cite{Kok.2008}, as their upper limit of $1\times10^8$, before equinox, was too small to allow for ice formation even under the rapid depletion scenario.

The most recent hypothesis for C$_4$N$_2$ ice formation in Titan's atmosphere was presented in \cite{Anderson.2016} where, rather than C$_4$N$_2$ ice forming directly from condensation, it is formed completely via solid-state photochemistry within HCN-HC$_3$N ice mixtures. This proposed mechanism, however, requires a column abundance of $10^{17}$ molecules/cm$^2$ to obtain a good spectral fit which is larger than the column abundance for cyanogen (C$_2$N$_2$) in photochemical model predictions \citep{Loison.2015}. Updated estimations of the upper limits on the gas phase abundance of C$_4$N$_2$ will help constrain the feasibility of C$_4$N$_2$ ice formation in Titan's atmosphere.

In the present study, we use high-resolution ($R \sim 90,000$) mid-IR spectroscopy with observations made by the Echelon-Cross-Echelle Spectrograph (EXES) aboard the SOFIA (Stratospheric Observatory Infrared Astronomy) aircraft at two spectral regions, 621 cm$^{-1}$ and 472 cm$^{-1}$, to search for C$_6$H$_2$ and C$_4$N$_2$ respectively. We perform radiative transfer models of the measured spectra to retrieve the atmospheric abundances of  C$_{4}$H$_{2}$ and C$_{3}$H$_{4}$ and derive upper limits for the undetected species, C$_6$H$_2$ and C$_4$N$_2$. We then compare these upper limits to photochemical model predictions of the target species and discuss the implications these upper limits have on our understanding of Titan's atmospheric chemistry. 

\begin{figure*}[ht!]
    \centering
    \includegraphics[width=0.9\textwidth]{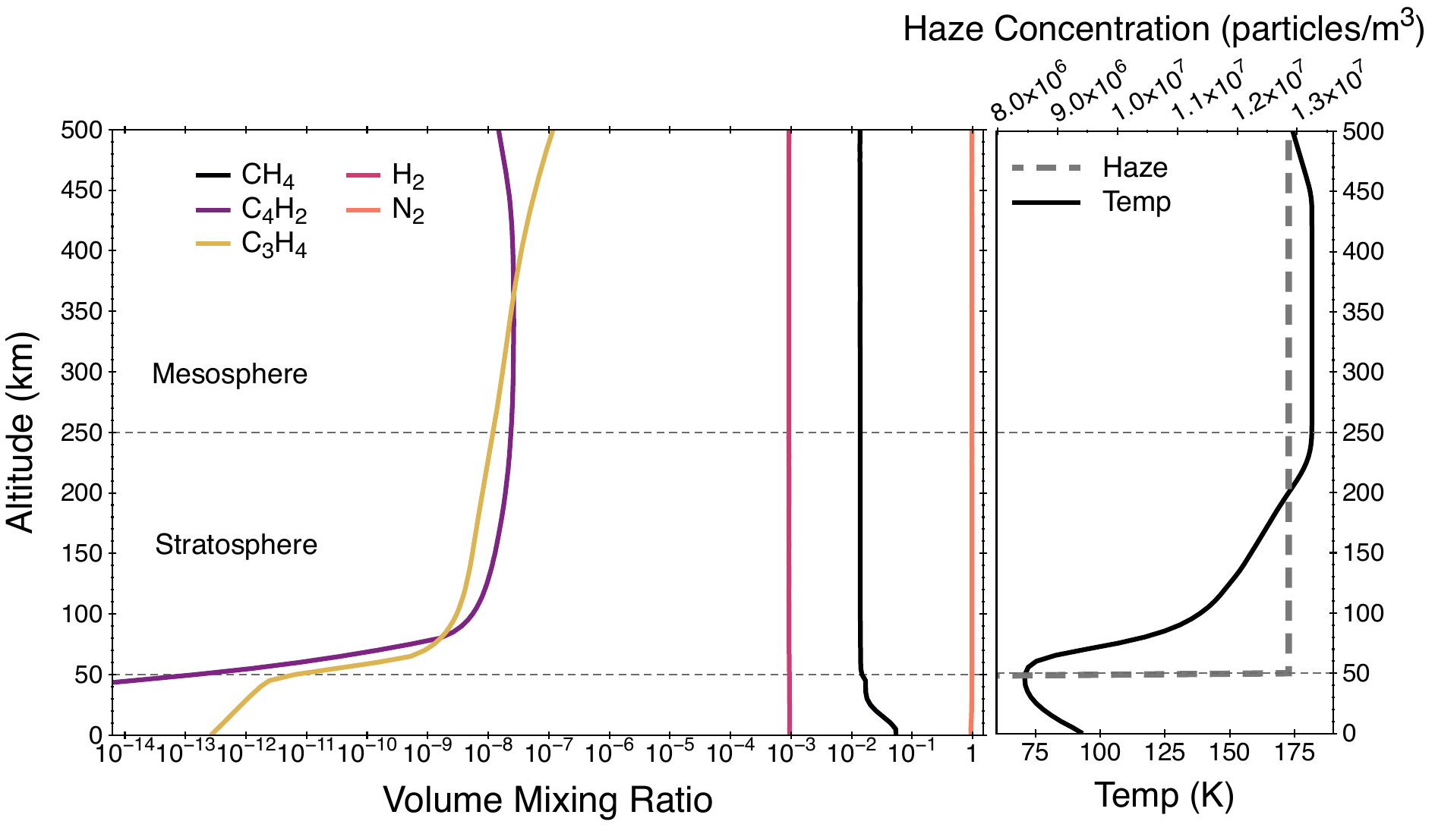}
    \caption{Reference profiles of the atmospheric gas composition (left panel), vertical temperature profile (right panel, solid black trace), and haze profiles (gray dashed trace). Temperature and gas profiles are from the photochemical model detailed in \cite{Vuitton.2019}.}
    \label{fig:apr_profs}
\end{figure*}

\section{EXES Observations} \label{sec:methods}

Observations of Titan's atmospheric spectra were made using EXES aboard NASA's SOFIA (Stratospheric Observatory for Infrared Astronomy) aircraft \citep{Richter.2018}. Details of the observations are listed in Table \ref{tab:obs}. Spectra were collected during flights 744 and 746 in June of 2021. EXES was operated in a high-medium configuration to obtain high-resolution ($R \sim 75000 - 90000$) spectra centered at 472 cm$^{-1}$ and 621 cm$^{-1}$ to observe the emission features of C$_4$N$_2$ for the former and C$_{4}$H$_{2}$, C$_{6}$H$_{2}$, and C$_{3}$H$_{4}$ for the latter region. Targets were nodded along the slit in an ABAB pattern by 7-8'' at intervals of 4 minutes to facilitate sky background subtraction.

The 2.65'' slit width was used for the 621 cm$^{-1}$ setting. Even at SOFIA’s altitude, the telluric spectrum at 621 cm$^{-1}$ has interference from densely spaced atmospheric lines of CO$_2$, O$_3$ and N$_2$O. To enable removal of these features, Callisto (the telluric calibrator) and Titan were observed on adjacent flight legs, at the same altitude and with an air mass difference of 0.16, ensuring the best possible match in telluric transmission. To accurately model the instrument line-shape, we derived the resolving power of the measurement by fitting synthetic transmission models to the Callisto spectrum which were generated by Planetary Spectrum Generator (PSG, \cite{Villanueva.2018}). We found the resolving power at the 621 cm$^{-1}$ spectral setting to be $R = 90000 \pm 2500$.

The 472 cm$^{-1}$ observations were taken on 3 flights with a slit width 3.2”. Due to the low density of telluric atmosphere features near 472 cm$^{-1}$, telluric interference was a minor issue that could be addressed by synthetic PSG atmosphere models and masking the regions containing features due to water. The telluric calibrator, Vesta was observed for this setting to mitigate risk, but it was only used to derive a resolving power for the 472 cm$^{-1}$ setting, $R = 75000 \pm 2500$.

All data were reduced with the EXES Redux pipeline \citep{Clarke.2015} version 3.0.1.dev.7 with the customary steps including spike removal, flat fielding, rectification of the cross-dispersed orders, coadding of nod-pairs, and extraction to the 1-D using optimal PSF weighting. Following data reduction using the Redux pipeline, spectral data points were binned using Nyquist sampling to a $\Delta\bar{\nu} = 0.003$ in order to reduce the noise in the spectrum. 
\begin{figure*}[ht!]
    \includegraphics[width = 0.95\textwidth]{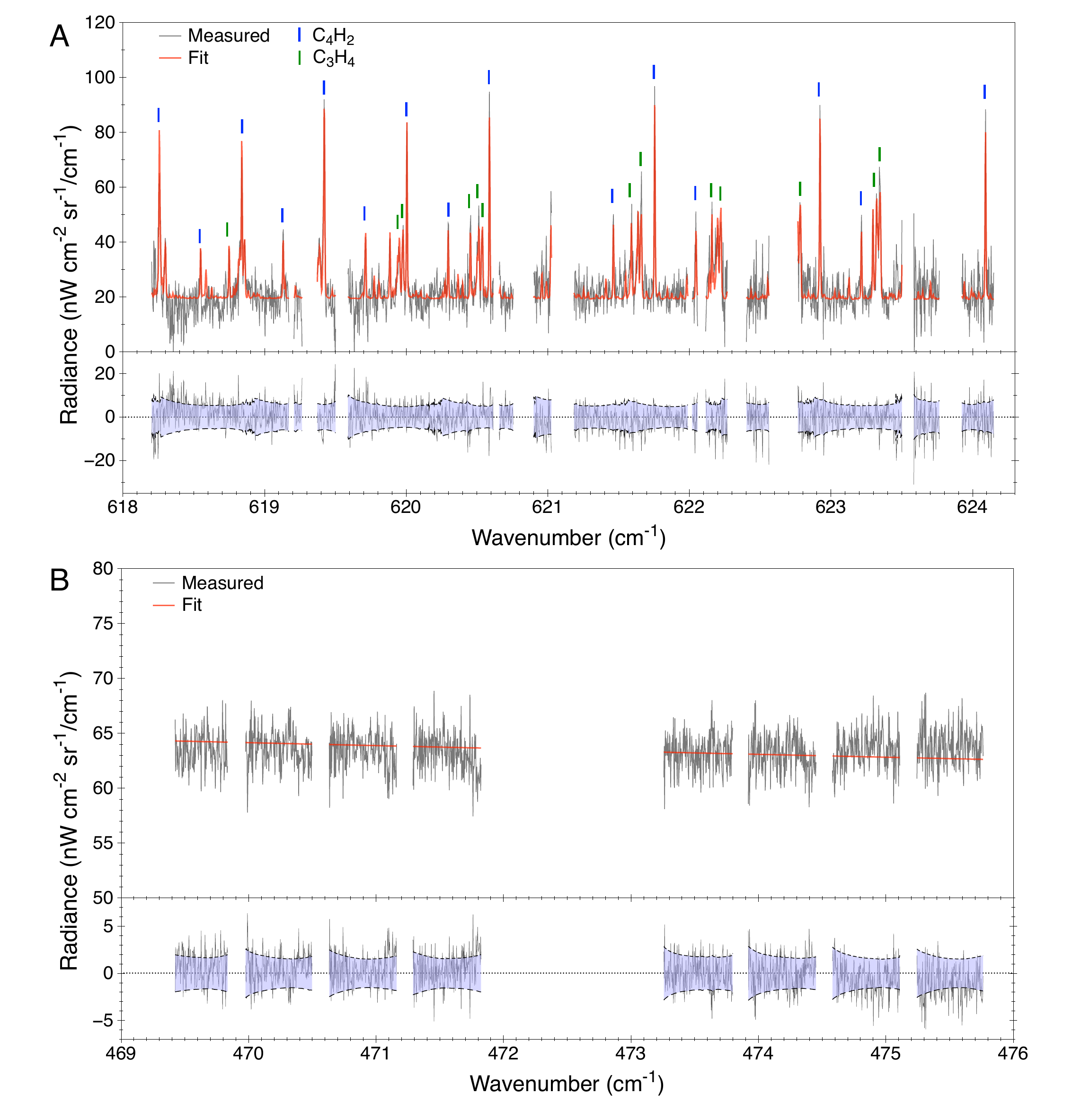}
    \caption{Model fit (red trace) to the measured spectrum (gray trace) for the 621 cm$^{-1}$ (A) and 472 cm$^{-1}$ (B) spectral settings. Bottom panels show the residual of the model fit to the measured spectra and a 1$\sigma$ estimation of the instrument uncertainty in the blue shaded regions. Emission features from C$_4$H$_2$ and C$_3$H$_4$ are indicated by the respective markers in A. There are no known emission features in the 472 cm$^{-1}$ spectral setting and only the haze and CIA continuum levels were fit in the retrieval.}
    \label{fig:specfit}
\end{figure*}

\section{Radiative Transfer Modeling}\label{sec:modeling}

Spectra were modeled using the archNEMESIS \citep{Alday.2025} radiative transfer model, which is a Python implementation of the Non-linear Optimal Estimator for Multivariate Spectral Analysis (NEMESIS) \citep{Irwin.2008}. NEMESIS has been used extensively to model Titan atmospheric spectra \citep{Nixon.2013,Teanby.2019,Wright.2024, Lombardo.2019,Thelen.2019} and has been used for modeling observations of Titan with TEXES (Texas Echelon-Cross-Echelle Spectrometer), the sister instrument to EXES \citep{Lombardo.2019}. archNEMESIS allows for both multi-core processing and larger spectral datasets due to improved memory efficiency in Python.

For this study, archNEMESIS is employed in two different ways to model the target species. First, the spectra are inverted using the optimal estimation method to retrieve the continuous atmospheric profiles of C$_{4}$H$_{2}$ and C$_{3}$H$_{4}$ (for the 621 cm$^{-1}$ spectral setting) as well as the scaling factor for Titan's haze profile (both spectral settings). We retrieved continuous profiles for the known gases (C$_{4}$H$_{2}$ and C$_{3}$H$_{4}$) and a scaling factor for the haze profile. We used an atmospheric model with 100 homogeneous layers equally spaced in log pressure to sufficiently sample the vertical profile of Titan's atmosphere. The a priori reference atmosphere used, shown in Figure \ref{fig:apr_profs}, uses gas and temperature profiles described in the \cite{Vuitton.2019} photochemical model. The photochemical model profiles for C$_{4}$H$_{2}$ and C$_{3}$H$_{4}$ were each scaled by 0.2 and 0.5, respectively, prior to retrieving the continuous profiles to ensure a good fit to the spectrum was achieved. To fit the continuum level, we retrieved a scaling factor for a uniform haze profile with an a priori concentration of $1.29 \times10^{7}$ particles/m$^{3}$ above 50 km (Figure \ref{fig:apr_profs}), using a uniform cross-section across the wavelength range for both spectra. 

Since the EXES observations are of Titan's disk, we modeled these observations using a weighted average of emission rays from different emission angles of the solid body and emission rays from the limb. We followed a similar approach to \cite{Teanby.2013} to both sufficiently sample Titan's radiance and calculate weights for each emission ray. In total, we used 55 emission rays from the disk center out to a limb height of 975 km.

\begin{figure}[ht!]
    \includegraphics[width = 0.48\textwidth]{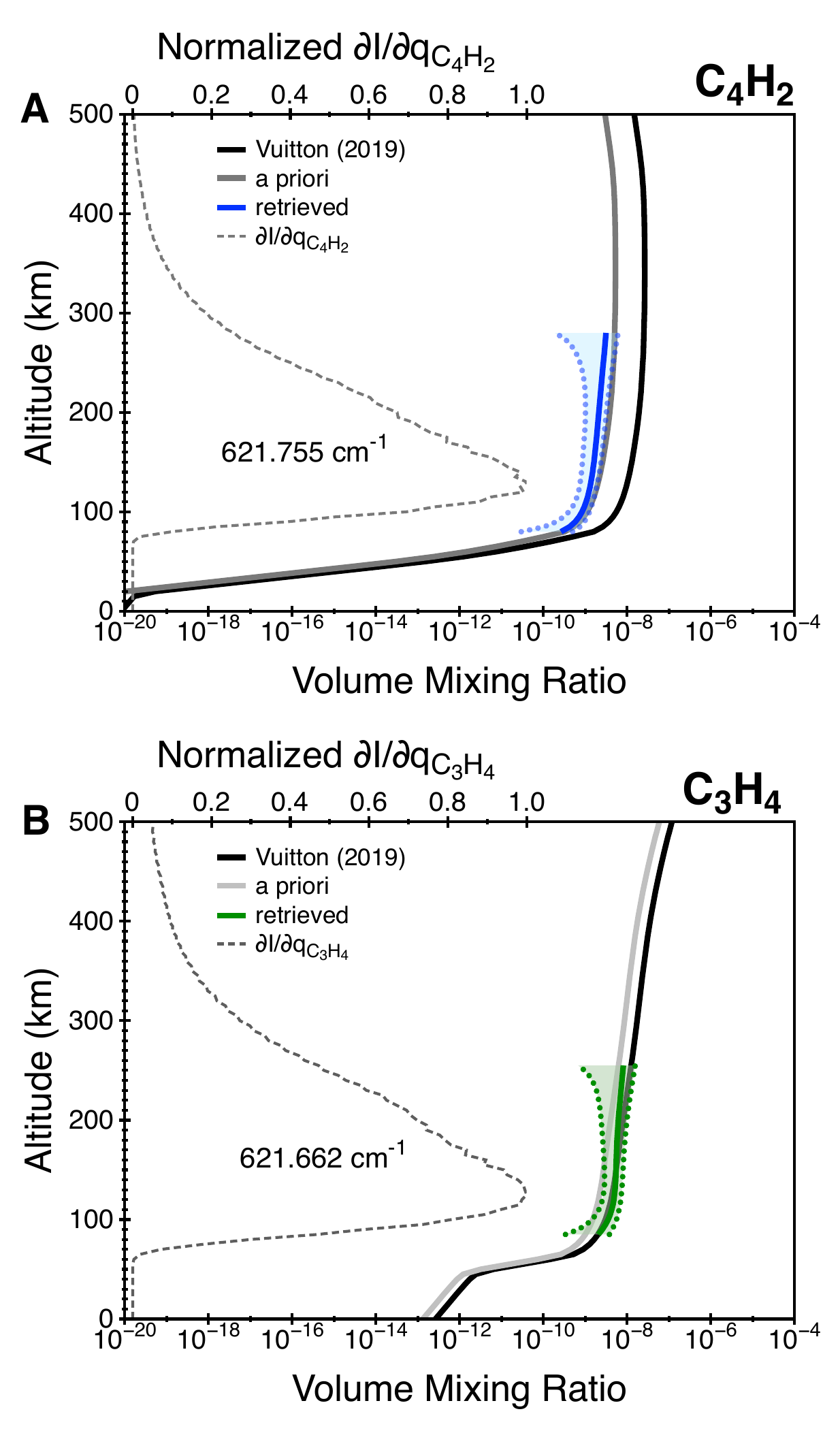}
    \caption{Photochemical model profiles (black trace) and retrieved profiles (colored trace) of C$_4$H$_2$ (A) and C$_3$H$_4$ (B). Each panel also shows the Jacobian at $\overline{\nu}_{max}$ for each molecule to indicate the altitude range the model is sensitive to.}
    \label{fig:contributions}
    
\end{figure}
\subsection{Spectroscopic Data}

Spectra were calculated in line-by-line mode using pre-tabulated line-by-line lookup tables calculated at 30 pressure and temperature levels between $5.6\times10^{-9}$ - $2.5$ bar and 70-250 K. Line data was then interpolated to the pressures and temperatures at each atmospheric layer in the atmospheric model. Spectra were also calculated using a Gaussian line shape at the resolving power of each spectral setting. The spectroscopic line lists for the known gas emission features at the 621 cm$^{-1}$ spectral setting, namely the $\nu_8$ band of C$_{4}$H$_{2}$ and the $\nu_9$ band of C$_{3}$H$_{4}$, were obtained from the HITRAN \citep{Gordon.2022} and GEISA \citep{Delahaye.2021} spectroscopic databases, respectively. The target species for this study, C$_{6}$H$_{2}$ and C$_{4}$N$_{2}$, are less commonly measured in atmospheric spectra and are not available in the standard databases. \cite{Shindo.2003} measured the mid-infrared spectrum of C$_{6}$H$_{2}$ and is the most up-to-date line-list for the $\nu_{11}$ band at $\sim 622$ cm$^{-1}$. Similarly, \cite{Jolly.2015} obtained the line list for the $\nu_8$ bending mode of C$_{4}$N$_{2}$ in a previous investigation of the upper limits of the gas phase species in Titan's atmosphere. Data for collision-induced absorption was obtained from the 2025 HITRAN database update \citep{Terragni.2025}.

\subsection{Determining Upper Limits}

Following the spectral fit, we determine the upper limits to the C$_6$H$_2$ and C$_4$N$_2$ abundance. We determine the goodness-of-fit for the modeled spectrum by summing the ratio of the square residual of the fit to the squared uncertainty of the measurement, described in Equation \ref{chi2}.
\begin{equation}
    \chi^2 = \frac{\Delta\bar{\nu}_{obs}}{\Delta \bar{\nu}_{res} } \sum_{i = 1}^{M}\frac{(I_{m}(\bar{\nu_i})-I_{r}(\bar{\nu_i}))^2}{\sigma_i^2}
    \label{chi2}
\end{equation}

Where $I_m$ and $I_r$ are the spectral radiance for the measurement and retrieved spectrum, respectively, at wavenumber $\bar{\nu}_i$, and $\sigma_i$ is the wavenumber dependent uncertainty. Once the emission features of the spectral region are well fit, the remaining residual will be due to instrument noise and any unfit gas emission features that were not included in the retrieval. The vertical profile of the target species is then set to be uniform volume mixing ratio (VMR) above a condensation altitude and we iterate over a range of VMR values, calculating a forward model of the spectrum at each abundance and determine the change in the $\chi^2$, $\Delta\chi^2$, with increased abundance of the target species. If the gas is present in detectable amounts, $\Delta\chi^2$ will decrease significantly, indicating the spectral fit is improved with the inclusion of the target species. Alternatively, when $\Delta\chi^2$ increases, the upper limits can be determined at $\Delta\chi^2 =  +1,+4$, and $+9$ for $1\sigma$, $2\sigma$, and $3\sigma$ upper limits, respectively. This method has been used in several previous studies of Titan's atmospheric trace constituents for detections and upper limits \citep{Nixon.2010, Teanby.2009, Jolly.2015}.

\section{Results}\label{sec:res}

\subsection{Abundance Retrievals for C$_4$H$_2$ and C$_3$H$_4$}

To fit the measured spectra, we used the optimal estimation method for retrieving the vertical profiles of the gases, C$_4$H$_2$ and C$_3$H$_4$, known to have emission features in the 621 cm$^{-1}$ spectral setting and a scaling factor for a uniform haze vertical profile at both spectral settings. The model fit to both spectra are shown in Figure \ref{fig:specfit}. We obtained a good fit ($\chi^2/N = 0.999$) to the measured 621 cm$^{-1}$ spectrum by fitting to the gas emission features of C$_4$H$_2$ and C$_3$H$_4$. The residual of the fit to the measurement is shown in the lower panel and is within the 1$\sigma$ estimation of the instrument noise. At lower wavenumbers, some of the emission features are slightly over fit and at longer wavenumbers the radiance of the stronger emission features are not fully captured, however the overall fit is good. C$_4$H$_2$ exhibits the strongest peaks, indicated by the blue vertical lines, with emission features originating from the P-branch of the $\nu_8$ vibrational mode. The emission features of C$_3$H$_4$, originating from the $\nu_9$ vibrational mode, are less intense, but tightly bunched and often near emission features from C$_4$H$_2$. Previously, these individual emission lines would not have been resolved with lower resolution spectrometers.

\begin{figure*}[ht!]
    \includegraphics[width = 0.9\textwidth]{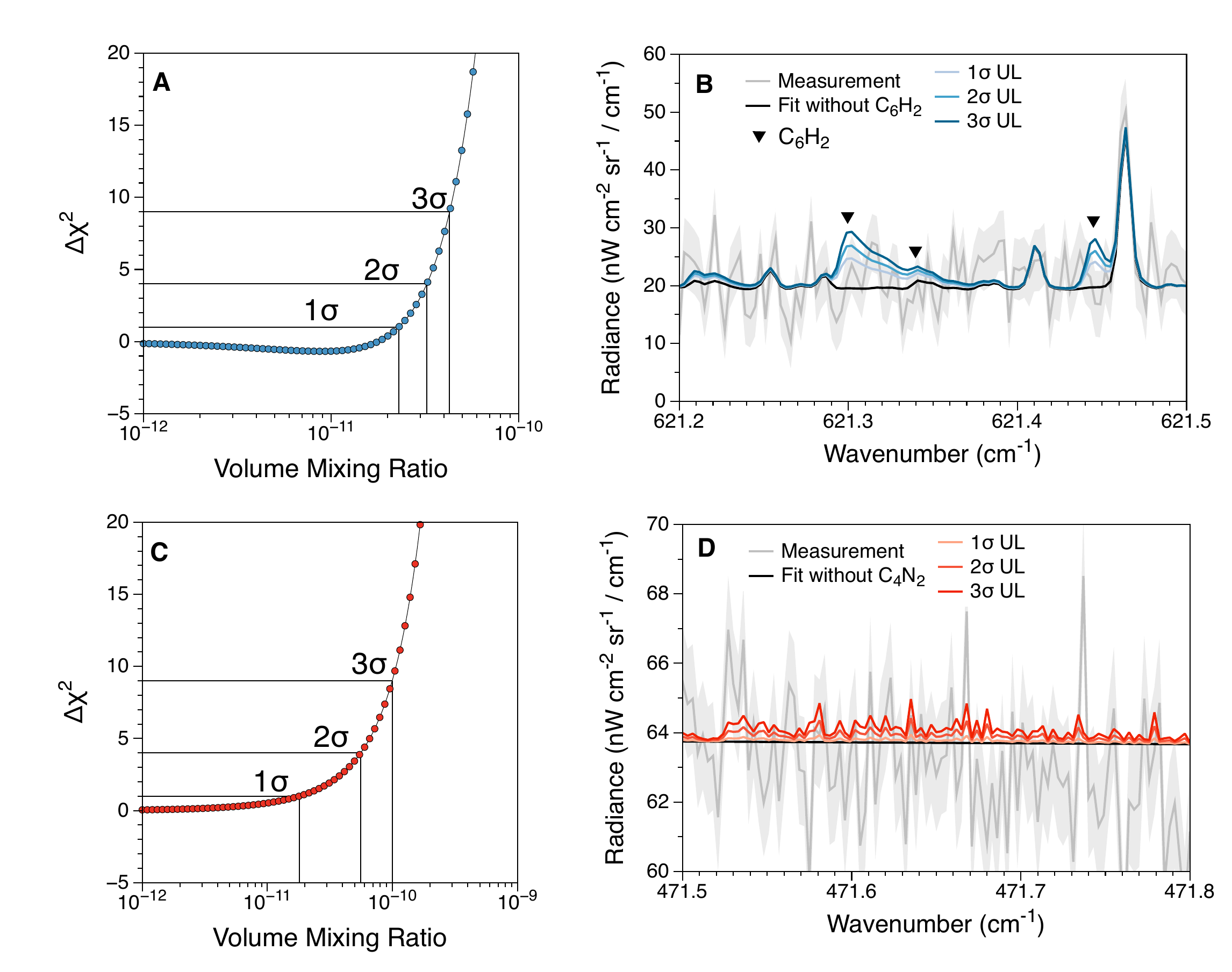}
    \caption{(A) The $\Delta\chi^2$ plotted as a function of  the VMR for the uniform profile of C$_6$H$_2$ and (B) the corresponding simulated spectra for the $1\sigma$, $2\sigma$, and $3\sigma$ upper limits with C$_6$H$_2$ peaks indicated by the black triangle markers. (C-D) The $\Delta\chi^2$ and corresponding spectra for the upper limits of  C$_4$N$_2$. All emission features in (D) are due to C$_4$N$_2$.}
    \label{fig:UL}
\end{figure*}

 Figure \ref{fig:contributions} shows the retrieved continuous profiles of each fit species in the altitude region where the Jacobian is peaked. Overall, the shape of the profiles was retained in the retrieval. For C$_3$H$_4$, there was slight enhancement in the stratosphere of the VMR bringing the abundance closer to the photochemical model prediction. For each species, the model was most sensitive in the stratosphere, where the majority of the gas abundance is found, peaking at $\sim130$ km ($\sim 5\times10^{-3}$ bar) for each species. At the 472 cm$^{-1}$ spectral setting, we only modeled the continuum level of the spectrum as no clear emission features were present. Without fitting any gas emission features, the model produced a $\chi^2/N = 1.01$, indicating the spectrum does not contain any significant features that are not being fit by the model.

\begin{deluxetable*}{cccccc}[ht!]
    \tablecaption{Upper limits for C$_6$H$_2$ and C$_4$N$_2$.}
    \label{UL}
    
    \tablehead{& & \multicolumn{3}{c}{Volume Mixing Ratio} & {Column Density (m$^{-2}$)}\\
        \colhead{Species} &\colhead{$1\sigma$NESR (nW cm$^{-2}$ sr$^{-1}$/cm$^{-1}$)}  & \colhead{$1\sigma$} & \colhead{$2\sigma$} & \colhead{$3\sigma$} & \colhead{$3\sigma$ }
    }
    \startdata
    C$_6$H$_2$ & 1.028 &  $<2.30\times10^{-11}$ & $<3.24\times10^{-11}$ & $<4.29\times10^{-11}$ & $<1.69\times10^{14}$\\
    C$_4$N$_2$ & 1.003 &  $<1.80\times10^{-11}$ & $<5.58\times10^{-11}$ & $<9.96\times10^{-11}$ & $<1.94\times10^{14}$\\
    \enddata
    \label{UL}
\end{deluxetable*}

\subsection{Upper limits for C$_6$H$_2$ and C$_4$N$_2$}

Following the retrievals, to determine the upper limits of C$_6$H$_2$ and C$_4$N$_2$ we simulate spectra with increasing abundances of the target species while keeping the retrieved profiles of the known gases and haze fixed. The difference between the $\chi^2$ of the updated spectrum and the retrieved spectrum ($\Delta\chi^2$) is calculated at each abundance or scale factor. If the target species is present, the $\Delta\chi^2$ value will decrease significantly, indicating the spectrum is better fit. A rejection of the detection occurs when the $\Delta\chi^2$ increases sharply with increased abundance. Figure \ref{fig:UL} shows the $\Delta\chi^2$ with increased VMR for C$_6$H$_2$ (A) and C$_4$N$_2$ (C). The cutoff altitude for C$_6$H$_2$ was set to 100 km, similarly to \cite{Shindo.2003}. The $\Delta\chi^2$ dips slightly below zero with increased abundance of C$_6$H$_2$, but is nowhere near a significance level to indicate detection, and increases sharply as the VMR increases above $10^{-11}$. The upper limit values for both C$_6$H$_2$ and C$_4$N$_2$ are reported in Table \ref{UL}. We find a $3\sigma$ upper limit on the uniform abundance of $4.29\times10^{-11}$ for C$_6$H$_2$. Figure \ref{fig:UL}(B) shows the corresponding spectra from each abundance at $1\sigma$, $2\sigma$, and $3\sigma$ upper limits. There is a clear growth in the radiance due to the $\nu_{11}$ vibrational mode of C$_6$H$_2$, indicated by the black triangle markers. The three main emission features in this spectra lie near strong emission features from C$_4$H$_2$, highlighting the importance of using the high-resolution capabilities of EXES. These emission features from C$_6$H$_2$ would have been smoothed together with emission features from C$_4$H$_2$ with lower resolution spectrometers such as CIRS or IRIS, requiring a much higher abundance of C$_6$H$_2$ to deviate significantly from the measurement and consequentially leading to higher upper limits.

A uniform vertical profile above 125 km was used to determine the upper limits to the abundance of gas phase C$_4$N$_2$. Figure \ref{fig:UL} (C) shows the $\Delta\chi^2$ increases sharply with VMR. We determine a $3\sigma$ upper limit on the abundance of $9.96\times10^{-11}$. The $\nu_{8}$ bending mode of C$_4$N$_2$ is very dense with many emission lines. Figure \ref{fig:UL}(D) shows the many emission  features due to C$_4$N$_2$ in the simulated spectra at each upper limit abundance. This high density of emission features provides strong statistical evidence for the rejection of C$_4$N$_2$ presence in the measurement, as each peak in the simulated spectra contributes to the total $\chi^2$ of the model.

\section{Discussion}\label{sec:disc}
\begin{figure*}[ht!]
    \centering
    \includegraphics[width = 0.9\textwidth]{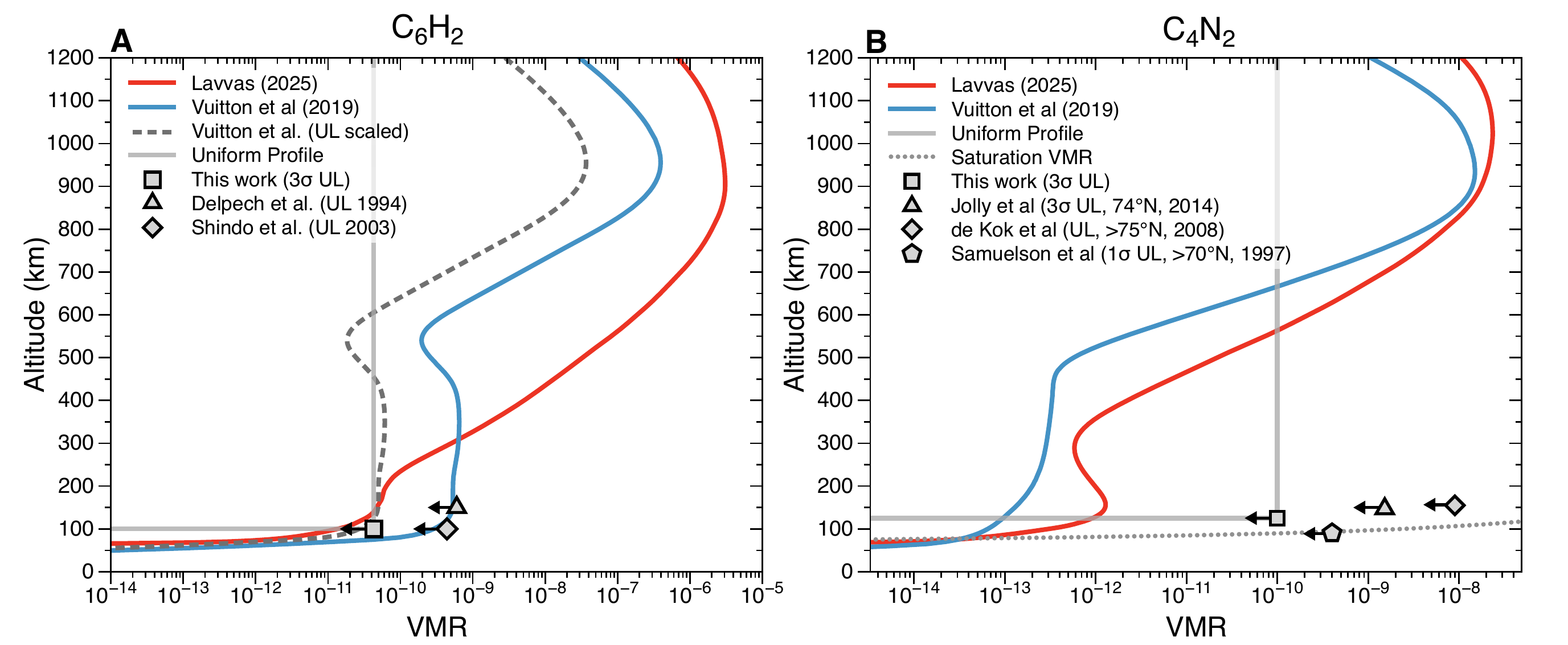}
    \caption{The 3$\sigma$ upper limits found for this study (gray squares) for C$_6$H$_2$ (A) and C$_4$N$_2$ (B) along with previous studies of the upper limits. Also included are the photochemical model vertical profiles of each species (red and blue traces), the uniform profiles used for the upper limits (gray solid trace), and the scaling factor for the \cite{Vuitton.2019} C$_6$H$_2$ profile (gray dashed trace). The gray dotted line in panel (B) is the saturation altitude calculated using the \cite{Vuitton.2019} temperature profile and equation for the saturation vapor pressures in \cite{Fray.2009}.}
    \label{fig:model comparison}
\end{figure*}
\subsection{C$_6$H$_2$}

Previous attempts at detecting C$_6$H$_2$ in Titan's atmosphere have ultimately led to non-detection and upper limit determinations of the abundance. Figure \ref{fig:model comparison} shows the upper limits determined from this study in comparison with those from \cite{Delpech.1994} and \cite{Shindo.2003}. The new upper limits improve upon the previous ones by over an order of magnitude. This can be attributed to the high-resolution of EXES which can resolve individual emission features in a small wavenumber range that may have previously been blended together at lower resolutions. Figure \ref{fig:model comparison} also compares the present results with photochemical model predictions of the vertical C$_6$H$_2$ profile. Comparing to the \cite{Vuitton.2019} profile of the C$_6$H$_2$ abundance, these upper limits are significantly lower than the predicted stratospheric abundance. We derived a scale factor of the \cite{Vuitton.2019} profile of 0.093.

Included in Figure \ref{fig:model comparison} is a vertical profile from an updated photochemical model for both C$_6$H$_2$ and C$_4$N$_2$ shared in private communication by Panayotis Lavvas (Lavvas (2025) in Figure \ref{fig:model comparison}). These profiles are the results of a  model similar to that presented in \cite{Vuitton.2019}, which is a 1D photochemical model of Titan's atmosphere including neutral and ionic chemical processes \citep{Lavvas.2008a,Lavvas.2021}. In addition to the model described in \cite{Lavvas.2008a}, the model includes high-resolution description of the energy deposition in Titan's upper atmosphere through solar photons \citep{Lavvas.2011}, and the contribution of the photochemical haze opacity in the radiation transfer in Titan's atmosphere based on measurements made by the Hyugens Descent Imager/Spectral Radiometer \citep{lavvas.2010}. The photochemical model also accounts for the escape of atomic and molecular hydrogen according to the Jean escape rate and methane is allowed to escape at a fixed velocity of 100 cm/s to match the observed profiles of methane and argon in the upper atmosphere based on measurements made by the Cassini Ion Neutral Mass Spectrometer \citep{Yelle.2008,Lavvas.2008b}.

The low upper limit for C$_6$H$_2$ in this work, combined with a lower retrieved abundance of its precursor, C$_4$H$_2$ suggests that loss mechanisms for C$_4$H$_2$ are not being fully represented in \cite{Vuitton.2019}. C$_4$H$_2$ can be lost in Titan's atmosphere through combination with atomic hydrogen (H + C$_4$H$_2$) \citep{Vuitton.2012, Vuitton.2019}. The Lavvas (2025) model prediction presented here does not include a heterogeneous loss mechanism of H to haze particle surfaces, described in \cite{Sekine.2008}, enriching the H abundance and decreasing the C$_4$H$_2$ abundance compared to \cite{Vuitton.2019}. Consequentially, this leads to a lower predicted C$_6$H$_2$ abundance as shown in Figure \ref{fig:model comparison}, which is in closer agreement with the present upper limits. Furthermore, additional loss of C$_4$H$_2$ could occur through the reaction with 1,3-butadiene (C$_4$H$_6$ isomer). \cite{Medvedkov.2025} recently showed through crossed molecular beam experiments that the C$_4$H radical can add directly to 1,3-butadiene or 2-methyl-1,3-butadiene followed by isomerization to form aromatic rings. Additionally, C$_2$H$_2$ has been shown to react with HCO+ to initiate a cascade of chemical reactions ultimately leading to the formation of benzene \citep{Pentsak.2024}. HCO+ has not been detected in Titan's atmosphere but is present in models of Titan's oxygen cycle \citep{Horst.2008, Dobrijevic.2016}. Such reactions that form aromatics from C$_2$H$_2$ and  C$_4$H$_2$, have not been well studied under Titan relevant conditions, but represent other potential loss mechanisms for the precursors to polyyne propagation in Titan's atmosphere. This suggests that polyyne propagation is not a suitable pathway to larger hydrocarbons or organic aerosols in Titan's atmosphere.

\subsection{C$_4$N$_2$}
The assignment of the emission feature at 478 cm$^{-1}$ to ice-phase C$_4$N$_2$ in both Voyager's IRIS data\citep{Samuelson.1997} and CIRS data \citep{Anderson.2016} has led to many studies attempting to understand the lack of gas phase C$_4$N$_2$ in spectra of Titan's atmosphere. Several upper-limits have been determined, including the most stringent ones from \cite{Jolly.2015} where a $3\sigma$ upper limit on the stratospheric abundance of $1.5\times10^{-9}$ using both nadir and limb observations of Titan's atmosphere using CIRS. Figure \ref{fig:model comparison}B shows the 3$\sigma$ upper limit presented in this study (gray square) in comparison with previous estimations of the C$_4$N$_2$ gas phase upper limits (gray points). The upper limits of this study lower the upper limits of the gas phase C$_4$N$_2$ abundance by an order of magnitude. This improvement largely stems from the high-resolution of EXES where the individual emission features of the C$_4$N$_2$ $\nu_8$ vibrational mode can be resolved. In previous studies of the upper limits, the lower resolution of CIRS and IRIS smooth the entire band into one peak, dampening the contribution of each line to the $\Delta\chi^2$. Figure \ref{fig:model comparison}B also compares the upper limits here to the photochemical model of \cite{Vuitton.2019} and the aforementioned Lavvas (2025) model. The photochemical models predict an abundance of C$_4$N$_2$ gas in the stratosphere between $\sim10^{-13} - 10^{-12}$ which is substantially lower than the upper limits presented here. 

Also present in Figure \ref{fig:model comparison}B are the altitudes at which C$_4$N$_2$ gas should condense (gray dotted line) based on the saturation vapor pressure temperatures presented in \cite{Fray.2009}. We calculated the saturation mixing ratios based on the temperature profile used in this study \citep{Vuitton.2019} based on the empirical determination of the saturation vapor pressure of C$_4$N$_2$ gas in \cite{Fray.2009}. Because the temperature changes dramatically with altitude in the stratosphere, the saturation mixing ratio also increases rapidly with altitude, hence the flat nature of the trace. We find that the upper limits presented here, calculated at 125 km, are below the saturation mixing ratio of C$_4$N$_2$ at this altitude which suggests that ice nucleation is not likely. 

Additionally, this value for the upper limit was determined from a disk averaged measurement, which is more sensitive to changes in the abundance at the equator where trace gas concentrations in Titan's atmosphere tend to be diminished \citep{Teanby.2008b, Vinatier.2020}. If C$_4$N$_2$ is present in lower abundances, it is possible, still that it could be enriched in the polar regions of titan during winter seasons, however, observations from Earth are obstructed from the winter poles. The current proposed mechanism of C$_4$N$_2$ formation is that C$_4$N$_2$ ice is formed totally in the solid-state through UV induced photochemistry \citep{Anderson.2016}. Based on this mechanism, \cite{Anderson.2016} requires a C$_4$N$_2$ column abundance of $10^{17}$ molecules/cm$^2$ for C$_4$N$_2$ which is substantially larger than the present value of $\sim2\times10^{14}$. This would suggest that the C$_4$N$_2$ ice is highly out of equilibrium with the gas phase and would need to be retained completely in the closed shell. Laboratory experiments studying more closely the gas and ice phase partitioning of C$_4$N$_2$ ice formed in closed shells would be needed to better understand the feasibility of this disequilibrium.

\section{Conclusions}\label{sec:conc}

This study presents spectra collected in the mid-infrared of Titan's atmosphere using EXES during the SOFIA mission. These spectra represent the lone study of Titan's atmosphere during the SOFIA mission but display the value of using high-resolution infrared spectroscopy for atmospheric studies. The spectra presented in this study showed spectral signatures for  C$_4$H$_2$ and C$_3$H$_4$, which allowed for their continuous stratospheric profiles to be retrieved. We showed the C$_3$H$_4$ profile to be similar to that of \cite{Vuitton.2019}, but the C$_4$H$_2$ profile was lower than the photochemical model prediction. The target species, C$_6$H$_2$ and C$_4$N$_2$ were not detected; instead, we performed an upper limits study to the atmospheric abundances of the target species. The C$_6$H$_2$ upper limit, which has not been assessed since the Voyager mission \citep{Delpech.1994}, was found to be lower than previous studies of \cite{Delpech.1994} and \cite{Shindo.2003} and constrains the photochemical model in \cite{Vuitton.2019}. We show by comparing to an update to the \cite{Lavvas.2008a} photochemical model, that C$_4$H$_2$ is over predicted in the \cite{Vuitton.2019} model and ultimately leads to the over prediction of the C$_6$H$_2$ stratospheric abundance. These results show the potential for C$_6$H$_2$ formation and large polyyne propagation is low and the precursors are likely lost to other chemical processes.

Additionally we lower the upper limit for C$_4$N$_2$ gas in Titan's atmosphere by an order of magnitude. This new value for the C$_4$N$_2$ upper limit shows that ice nucleation is unlikely to occur in Titan's atmosphere directly. The formation of C$_4$N$_2$ ice through solid-state photochemistry is still possible, however, the ice would be substantially out of equilibrium with the gas phase. Further work would be needed to better understand the gas-solid phase partitioning of C$_4$N$_2$ ice formed in an enclosed shell.

\section*{Acknowledgments}
The material is based upon work supported by NASA under award number 80GSFC24M0006. NAT is supported by UK Science and Technology Facilities Council grant ST/Y000676/1.

%

\vspace{5mm}
\facilities{SOFIA, EXES}


\software{archNEMESIS
          }




\bibliography{References}{}

\begin{thebibliography}{}
\expandafter\ifx\csname natexlab\endcsname\relax\def\natexlab#1{#1}\fi
\providecommand{\url}[1]{\href{#1}{#1}}
\providecommand{\dodoi}[1]{doi:~\href{http://doi.org/#1}{\nolinkurl{#1}}}
\providecommand{\doeprint}[1]{\href{http://ascl.net/#1}{\nolinkurl{http://ascl.net/#1}}}
\providecommand{\doarXiv}[1]{\href{https://arxiv.org/abs/#1}{\nolinkurl{https://arxiv.org/abs/#1}}}

\bibitem[{Alday {et~al.}(2025)Alday, Penn, Irwin, Mason, Yang, \& Dobinson}]{Alday.2025}
Alday, J., Penn, J., Irwin, P., {et~al.} 2025, Journal of Open Research Software, 13, \dodoi{10.5334/jors.554}

\bibitem[{Anderson {et~al.}(2016)Anderson, Samuelson, Yung, \& McLain}]{Anderson.2016}
Anderson, C.~M., Samuelson, R.~E., Yung, Y.~L., \& McLain, J.~L. 2016, Geophysical Research Letters, 43, 3088, \dodoi{10.1002/2016gl067795}

\bibitem[{Cernicharo {et~al.}(2001{\natexlab{a}})Cernicharo, Heras, Pardo, Tielens, Guélin, Dartois, Neri, \& Waters}]{Cernicharo.2001b}
Cernicharo, J., Heras, A.~M., Pardo, J.~R., {et~al.} 2001{\natexlab{a}}, The Astrophysical Journal Letters, 546, L127–L130, \dodoi{10.1086/318872}

\bibitem[{Cernicharo {et~al.}(2001{\natexlab{b}})Cernicharo, Heras, Tielens, Pardo, Herpin, Guélin, \& Waters}]{Cernicharo.2001a}
Cernicharo, J., Heras, A.~M., Tielens, A. G. G.~M., {et~al.} 2001{\natexlab{b}}, The Astrophysical Journal Letters, 546, L123–L126, \dodoi{10.1086/318871}

\bibitem[{Chassefi{\`e}re \& Cabane(1995)}]{Chassefiere.1995}
Chassefi{\`e}re, E., \& Cabane, M. 1995, Planetary and Space Science, 43, 91–103, \dodoi{10.1016/0032-0633(94)00138-h}

\bibitem[{{Clarke} {et~al.}(2015){Clarke}, {Vacca}, \& {Shuping}}]{Clarke.2015}
{Clarke}, M., {Vacca}, W.~D., \& {Shuping}, R.~Y. 2015, in Astronomical Society of the Pacific Conference Series, Vol. 495, Astronomical Data Analysis Software an Systems XXIV (ADASS XXIV), ed. A.~R. {Taylor} \& E.~{Rosolowsky}, 355

\bibitem[{Coustenis {et~al.}(2003)Coustenis, Salama, Schulz, Ott, Lellouch, Encrenaz, Gautier, \& Feuchtgruber}]{Coustenis.2003}
Coustenis, A., Salama, A., Schulz, B., {et~al.} 2003, Icarus, 161, 383–403, \dodoi{10.1016/s0019-1035(02)00028-3}

\bibitem[{Coustenis {et~al.}(1998)Coustenis, Salama, Lellouch, Encrenaz, Bjoraker, Samuelson, Graauw, Feuchtgruber, \& Kessler}]{Coustenis.1998}
Coustenis, A., Salama, A., Lellouch, E., {et~al.} 1998, Astronomy and Astrophysics, L85–L89

\bibitem[{Coustenis {et~al.}(2010)Coustenis, Jennings, Nixon, Achterberg, Lavvas, Vinatier, Teanby, Bjoraker, Carlson, Piani, Bampasidis, Flasar, \& Romani}]{Coustenis.2010}
Coustenis, A., Jennings, D., Nixon, C., {et~al.} 2010, Icarus, 207, 461–476, \dodoi{10.1016/j.icarus.2009.11.027}

\bibitem[{de~Arag\~ao {et~al.}(2025)de~Arag\~ao, Liang, Mancini, Vanuzzo, Pannacci, Faginas-Lago, Casavecchia, Rosi, \& Balucani}]{Aragao.2025}
de~Arag\~ao, E. V. F.~d., Liang, P., Mancini, L., {et~al.} 2025, ACS Earth and Space Chemistry, 9, 2199–2214, \dodoi{10.1021/acsearthspacechem.5c00154}

\bibitem[{de~Kok {et~al.}(2008)de~Kok, Irwin, \& Teanby}]{Kok.2008}
de~Kok, R., Irwin, P., \& Teanby, N. 2008, Icarus, 197, 572, \dodoi{10.1016/j.icarus.2008.05.024}

\bibitem[{Delahaye {et~al.}(2021)Delahaye, Armante, Scott, Jacquinet-Husson, Chédin, Crépeau, Crevoisier, Douet, Perrin, Barbe, Boudon, Campargue, Coudert, Ebert, Flaud, Gamache, Jacquemart, Jolly, Tchana, Kyuberis, Li, Lyulin, Manceron, Mikhailenko, Moazzen-Ahmadi, Müller, Naumenko, Nikitin, Perevalov, Richard, Starikova, Tashkun, Tyuterev, Auwera, Vispoel, Yachmenev, \& Yurchenko}]{Delahaye.2021}
Delahaye, T., Armante, R., Scott, N., {et~al.} 2021, Journal of Molecular Spectroscopy, 380, 111510, \dodoi{10.1016/j.jms.2021.111510}

\bibitem[{Delpech {et~al.}(1994)Delpech, Guillemin, Paillous, Khlifi, Bruston, \& Raulin}]{Delpech.1994}
Delpech, C., Guillemin, J., Paillous, P., {et~al.} 1994, Spectrochimica Acta Part A: Molecular Spectroscopy, 50, 1095, \dodoi{10.1016/0584-8539(94)80031-6}

\bibitem[{Dobrijevic {et~al.}(2016)Dobrijevic, Loison, Hickson, \& Gronoff}]{Dobrijevic.2016}
Dobrijevic, M., Loison, J., Hickson, K., \& Gronoff, G. 2016, Icarus, 268, 313–339, \dodoi{10.1016/j.icarus.2015.12.045}

\bibitem[{Encrenaz(2003)}]{Encrenaz.2003}
Encrenaz, T. 2003, Planetary and Space Science, 51, 89–103, \dodoi{10.1016/s0032-0633(02)00145-9}

\bibitem[{Fray \& Schmitt(2009)}]{Fray.2009}
Fray, N., \& Schmitt, B. 2009, Planetary and Space Science, 57, 2053–2080, \dodoi{10.1016/j.pss.2009.09.011}

\bibitem[{Gordon {et~al.}(2022)Gordon, Rothman, Hargreaves, Hashemi, Karlovets, Skinner, Conway, Hill, Kochanov, Tan, Wcisło, Finenko, Nelson, Bernath, Birk, Boudon, Campargue, Chance, Coustenis, Drouin, Flaud, Gamache, Hodges, Jacquemart, Mlawer, Nikitin, Perevalov, Rotger, Tennyson, Toon, Tran, Tyuterev, Adkins, Baker, Barbe, Canè, Császár, Dudaryonok, Egorov, Fleisher, Fleurbaey, Foltynowicz, Furtenbacher, Harrison, Hartmann, Horneman, Huang, Karman, Karns, Kassi, Kleiner, Kofman, Kwabia–Tchana, Lavrentieva, Lee, Long, Lukashevskaya, Lyulin, Makhnev, Matt, Massie, Melosso, Mikhailenko, Mondelain, Müller, Naumenko, Perrin, Polyansky, Raddaoui, Raston, Reed, Rey, Richard, Tóbiás, Sadiek, Schwenke, Starikova, Sung, Tamassia, Tashkun, Auwera, Vasilenko, Vigasin, Villanueva, Vispoel, Wagner, Yachmenev, \& Yurchenko}]{Gordon.2022}
Gordon, I., Rothman, L., Hargreaves, R., {et~al.} 2022, Journal of Quantitative Spectroscopy and Radiative Transfer, 277, 107949, \dodoi{10.1016/j.jqsrt.2021.107949}

\bibitem[{Gu {et~al.}(2009)Gu, Kim, Kaiser, Mebel, Liang, \& Yung}]{Gu.2009}
Gu, X., Kim, Y.~S., Kaiser, R.~I., {et~al.} 2009, Proceedings of the National Academy of Sciences, 106, 16078, \dodoi{10.1073/pnas.0900525106}

\bibitem[{Hanel {et~al.}(1981)Hanel, Conrath, Flasar, Kunde, Maguire, Pearl, Pirraglia, Samuelson, Herath, Allison, Cruikshank, Gautier, Gierasch, Horn, Koppany, \& Ponnamperuma}]{Hanel.1981}
Hanel, R., Conrath, B., Flasar, F.~M., {et~al.} 1981, Science, 212, 192–200, \dodoi{10.1126/science.212.4491.192}

\bibitem[{H{\"o}rst(2017)}]{Horst.2017}
H{\"o}rst, S.~M. 2017, Journal of Geophysical Research: Planets, 122, 432–482, \dodoi{10.1002/2016je005240}

\bibitem[{H{\"o}rst {et~al.}(2008)H{\"o}rst, Vuitton, \& Yelle}]{Horst.2008}
H{\"o}rst, S.~M., Vuitton, V., \& Yelle, R.~V. 2008, Journal of Geophysical Research: Planets, 113, \dodoi{10.1029/2008je003135}

\bibitem[{Irwin {et~al.}(2008)Irwin, Teanby, Kok, Fletcher, Howett, Tsang, Wilson, Calcutt, Nixon, \& Parrish}]{Irwin.2008}
Irwin, P., Teanby, N., Kok, R.~d., {et~al.} 2008, Journal of Quantitative Spectroscopy and Radiative Transfer, 109, 1136, \dodoi{10.1016/j.jqsrt.2007.11.006}

\bibitem[{Jolly {et~al.}(2015)Jolly, Cottini, Fayt, Manceron, Kwabia-Tchana, Benilan, Guillemin, Nixon, \& Irwin}]{Jolly.2015}
Jolly, A., Cottini, V., Fayt, A., {et~al.} 2015, Icarus, 248, 340, \dodoi{10.1016/j.icarus.2014.10.049}

\bibitem[{Kuiper(1944)}]{Kuiper.1944}
Kuiper, G.~P. 1944, Astrophysical Journal, 378–383

\bibitem[{Kunde {et~al.}(1981)Kunde, Aikin, Hanel, Jennings, Maguire, \& Samuelson}]{Kunde.1981}
Kunde, V.~G., Aikin, A.~C., Hanel, R.~A., {et~al.} 1981, Nature, 292, 686, \dodoi{10.1038/292686a0}

\bibitem[{Lavvas {et~al.}(2008{\natexlab{a}})Lavvas, Coustenis, \& Vardavas}]{Lavvas.2008a}
Lavvas, P., Coustenis, A., \& Vardavas, I. 2008{\natexlab{a}}, Planetary and Space Science, 56, 27–66, \dodoi{10.1016/j.pss.2007.05.026}

\bibitem[{Lavvas {et~al.}(2008{\natexlab{b}})Lavvas, Coustenis, \& Vardavas}]{Lavvas.2008b}
---. 2008{\natexlab{b}}, Planetary and Space Science, 56, 67–99, \dodoi{10.1016/j.pss.2007.05.027}

\bibitem[{Lavvas {et~al.}(2011)Lavvas, Galand, Yelle, Heays, Lewis, Lewis, \& Coates}]{Lavvas.2011}
Lavvas, P., Galand, M., Yelle, R., {et~al.} 2011, Icarus, 213, 233–251, \dodoi{10.1016/j.icarus.2011.03.001}

\bibitem[{Lavvas {et~al.}(2021)Lavvas, Lellouch, Strobel, Gurwell, Cheng, Young, \& Gladstone}]{Lavvas.2021}
Lavvas, P., Lellouch, E., Strobel, D.~F., {et~al.} 2021, Nature Astronomy, 5, 289–297, \dodoi{10.1038/s41550-020-01270-3}

\bibitem[{Lavvas {et~al.}(2010)Lavvas, Yelle, \& Griffith}]{lavvas.2010}
Lavvas, P., Yelle, R., \& Griffith, C. 2010, Icarus, 210, 832–842, \dodoi{10.1016/j.icarus.2010.07.025}

\bibitem[{Lebonnois {et~al.}(2002)Lebonnois, Bakes, \& McKay}]{Lebonnois.2002}
Lebonnois, S., Bakes, E., \& McKay, C.~P. 2002, Icarus, 159, 505–517, \dodoi{10.1006/icar.2002.6943}

\bibitem[{Letourneur \& Coustenis(1993)}]{Letourneur.1993}
Letourneur, B., \& Coustenis, A. 1993, Planetary and Space Science, 41, 593–602, \dodoi{10.1016/0032-0633(93)90079-h}

\bibitem[{Loison {et~al.}(2015)Loison, Hébrard, Dobrijevic, Hickson, Caralp, Hue, Gronoff, Venot, \& Bénilan}]{Loison.2015}
Loison, J., Hébrard, E., Dobrijevic, M., {et~al.} 2015, Icarus, 247, 218–247, \dodoi{10.1016/j.icarus.2014.09.039}

\bibitem[{Lombardo {et~al.}(2019)Lombardo, Nixon, Greathouse, Bézard, Jolly, Vinatier, Teanby, Richter, Irwin, Coustenis, \& Flasar}]{Lombardo.2019}
Lombardo, N.~A., Nixon, C.~A., Greathouse, T.~K., {et~al.} 2019, The Astrophysical Journal Letters, 881, L33, \dodoi{10.3847/2041-8213/ab3860}

\bibitem[{Maguire {et~al.}(1981)Maguire, Hanel, Jennings, Kunde, \& Samuelson}]{Maguire.1981}
Maguire, W.~C., Hanel, R.~A., Jennings, D.~E., Kunde, V.~G., \& Samuelson, R.~E. 1981, Nature, 292, 683–686, \dodoi{10.1038/292683a0}

\bibitem[{Medvedkov {et~al.}(2025)Medvedkov, Yang, Nikolayev, Goettl, Eckhardt, Mebel, \& Kaiser}]{Medvedkov.2025}
Medvedkov, I.~A., Yang, Z., Nikolayev, A.~A., {et~al.} 2025, The Journal of Physical Chemistry Letters, 16, 658, \dodoi{10.1021/acs.jpclett.4c03150}

\bibitem[{Nixon(2024)}]{Nixon.2024}
Nixon, C.~A. 2024, ACS Earth and Space Chemistry, 8, 406, \dodoi{10.1021/acsearthspacechem.2c00041}

\bibitem[{Nixon {et~al.}(2010)Nixon, Achterberg, Teanby, Irwin, Flaud, Kleiner, Dehayem-Kamadjeu, Brown, Sams, Bézard, Coustenis, Ansty, Mamoutkine, Vinatier, Bjoraker, Jennings, Romani, \& Flasar}]{Nixon.2010}
Nixon, C.~A., Achterberg, R.~K., Teanby, N.~A., {et~al.} 2010, Faraday Discussions, 147, 65, \dodoi{10.1039/c003771k}

\bibitem[{Nixon {et~al.}(2013)Nixon, Jennings, Bézard, Vinatier, Teanby, Sung, Ansty, Irwin, Gorius, Cottini, Coustenis, \& Flasar}]{Nixon.2013}
Nixon, C.~A., Jennings, D.~E., Bézard, B., {et~al.} 2013, The Astrophysical Journal Letters, 776, L14, \dodoi{10.1088/2041-8205/776/1/l14}

\bibitem[{Nixon {et~al.}(2019)Nixon, Ansty, Lombardo, Bjoraker, Achterberg, Annex, Rice, Romani, Jennings, Samuelson, Anderson, Coustenis, Bézard, Vinatier, Lellouch, Courtin, Teanby, Cottini, \& Flasar}]{Nixon.2019}
Nixon, C.~A., Ansty, T.~M., Lombardo, N.~A., {et~al.} 2019, The Astrophysical Journal Supplement Series, 244, 14, \dodoi{10.3847/1538-4365/ab3799}

\bibitem[{Nixon {et~al.}(2020)Nixon, Thelen, Cordiner, Kisiel, Charnley, Molter, Serigano, Irwin, Teanby, \& Kuan}]{Nixon.2020}
Nixon, C.~A., Thelen, A.~E., Cordiner, M.~A., {et~al.} 2020, The Astronomical Journal, 160, 205, \dodoi{10.3847/1538-3881/abb679}

\bibitem[{Nixon {et~al.}(2025)Nixon, B{\'e}zard, Cornet, Coy, Pater, Es-Sayeh, Hammel, Lellouch, Lombardo, L{\'o}pez-Puertas, Lora, Rannou, Rodriguez, Teanby, Turtle, Achterberg, Alvarez, Davies, Kleer, Doppmann, Fletcher, Hayes, Holler, Irwin, Jordan, King, Kutsop, Marlin, Melin, Milam, Molter, Moore, Nyffenegger-P{\'e}r{\'e}, O’Donoghue, O’Meara, Rafkin, Roman, Rostopchina, Rowe-Gurney, Schmidt, Schmidt, Sotin, Stallard, Stansberry, \& West}]{Nixon.2025}
Nixon, C.~A., B{\'e}zard, B., Cornet, T., {et~al.} 2025, Nature Astronomy, 9, 969–981, \dodoi{10.1038/s41550-025-02537-3}

\bibitem[{Palmer {et~al.}(2017)Palmer, Cordiner, Nixon, Charnley, Teanby, Kisiel, Irwin, \& Mumma}]{Palmer.2017}
Palmer, M.~Y., Cordiner, M.~A., Nixon, C.~A., {et~al.} 2017, Science Advances, 3, e1700022, \dodoi{10.1126/sciadv.1700022}

\bibitem[{Pentsak {et~al.}(2024)Pentsak, Murga, \& Ananikov}]{Pentsak.2024}
Pentsak, E.~O., Murga, M.~S., \& Ananikov, V.~P. 2024, ACS Earth and Space Chemistry, 8, 798–856, \dodoi{10.1021/acsearthspacechem.3c00223}

\bibitem[{Richter {et~al.}(2018)Richter, DeWitt, McKelvey, Montiel, McMurray, \& Case}]{Richter.2018}
Richter, M.~J., DeWitt, C.~N., McKelvey, M., {et~al.} 2018, Journal of Astronomical Instrumentation, 07, 1840013, \dodoi{10.1142/s2251171718400135}

\bibitem[{Samuelson {et~al.}(1997)Samuelson, Mayo, Knuckles, \& Khanna}]{Samuelson.1997}
Samuelson, R., Mayo, L., Knuckles, M., \& Khanna, R. 1997, Planetary and Space Science, 45, 941, \dodoi{10.1016/s0032-0633(97)00088-3}

\bibitem[{Samuelson {et~al.}(1983)Samuelson, Maguire, Hanel, Kunde, Jennings, Yung, \& Aikin}]{Samuelson.1983}
Samuelson, R.~E., Maguire, W.~C., Hanel, R.~A., {et~al.} 1983, Journal of Geophysical Research: Space Physics, 88, 8709–8715, \dodoi{10.1029/ja088ia11p08709}

\bibitem[{Sekine {et~al.}(2008)Sekine, Lebonnois, Imanaka, Matsui, Bakes, McKay, Khare, \& Sugita}]{Sekine.2008}
Sekine, Y., Lebonnois, S., Imanaka, H., {et~al.} 2008, Icarus, 194, 201–211, \dodoi{10.1016/j.icarus.2007.08.030}

\bibitem[{Shindo {et~al.}(2003)Shindo, Benilan, Guillemin, Chaquin, Jolly, \& Raulin}]{Shindo.2003}
Shindo, F., Benilan, Y., Guillemin, J.-C., {et~al.} 2003, Planetary and Space Science, 51, 9, \dodoi{10.1016/s0032-0633(02)00151-4}

\bibitem[{Teanby {et~al.}(2009)Teanby, Irwin, Kok, Jolly, Bézard, Nixon, \& Calcutt}]{Teanby.2009}
Teanby, N., Irwin, P., Kok, R.~d., {et~al.} 2009, Icarus, 202, 620, \dodoi{10.1016/j.icarus.2009.03.022}

\bibitem[{Teanby {et~al.}(2008{\natexlab{a}})Teanby, Irwin, Kok, Nixon, Coustenis, Royer, Calcutt, Bowles, Fletcher, Howett, \& Taylor}]{Teanby.2008a}
---. 2008{\natexlab{a}}, Icarus, 193, 595–611, \dodoi{10.1016/j.icarus.2007.08.017}

\bibitem[{Teanby {et~al.}(2013)Teanby, Irwin, Nixon, Courtin, Swinyard, Moreno, Lellouch, Rengel, \& Hartogh}]{Teanby.2013}
Teanby, N., Irwin, P., Nixon, C., {et~al.} 2013, Planetary and Space Science, 75, 136–147, \dodoi{10.1016/j.pss.2012.11.008}

\bibitem[{Teanby {et~al.}(2019)Teanby, Sylvestre, Sharkey, Nixon, Vinatier, \& Irwin}]{Teanby.2019}
Teanby, N.~A., Sylvestre, M., Sharkey, J., {et~al.} 2019, Geophysical Research Letters, 46, 3079–3089, \dodoi{10.1029/2018gl081401}

\bibitem[{Teanby {et~al.}(2008{\natexlab{b}})Teanby, Kok, Irwin, Osprey, Vinatier, Gierasch, Read, Flasar, Conrath, Achterberg, Bézard, Nixon, \& Calcutt}]{Teanby.2008b}
Teanby, N.~A., Kok, R.~d., Irwin, P. G.~J., {et~al.} 2008{\natexlab{b}}, Journal of Geophysical Research: Planets, 113, \dodoi{10.1029/2008je003218}

\bibitem[{Terragni {et~al.}(2025)Terragni, Gordon, Adkins, Boulet, Campargue, Chistikov, Finenko, Finkenzeller, Fleurbaey, Hargreaves, Hanson, Hartmann, Klingberg, Kohler, Koroleva, Mondelain, Piccioni, Stefani, Strand, Tran, Turbet, Vigasin, Vitali, Volkamer, \& Wei}]{Terragni.2025}
Terragni, J., Gordon, I., Adkins, E., {et~al.} 2025, Journal of Quantitative Spectroscopy and Radiative Transfer, 109631, \dodoi{10.1016/j.jqsrt.2025.109631}

\bibitem[{Thelen {et~al.}(2019)Thelen, Nixon, Cordiner, Charnley, Irwin, \& Kisiel}]{Thelen.2019}
Thelen, A.~E., Nixon, C.~A., Cordiner, M.~A., {et~al.} 2019, The Astronomical Journal, 157, 219, \dodoi{10.3847/1538-3881/ab19bb}

\bibitem[{Trainer {et~al.}(2006)Trainer, Pavlov, DeWitt, Jimenez, McKay, Toon, \& Tolbert}]{Trainer.2006}
Trainer, M.~G., Pavlov, A.~A., DeWitt, H.~L., {et~al.} 2006, Proceedings of the National Academy of Sciences, 103, 18035–18042, \dodoi{10.1073/pnas.0608561103}

\bibitem[{{Villanueva} {et~al.}(2018){Villanueva}, {Smith}, {Protopapa}, {Faggi}, \& {Mandell}}]{Villanueva.2018}
{Villanueva}, G.~L., {Smith}, M.~D., {Protopapa}, S., {Faggi}, S., \& {Mandell}, A.~M. 2018, \jqsrt, 217, 86, \dodoi{10.1016/j.jqsrt.2018.05.023}

\bibitem[{Vinatier {et~al.}(2020)Vinatier, Mathé, Bézard, d’Ollone, Lebonnois, Dauphin, Flasar, Achterberg, Seignovert, Sylvestre, Teanby, Gorius, Mamoutkine, Guandique, \& Jennings}]{Vinatier.2020}
Vinatier, S., Mathé, C., Bézard, B., {et~al.} 2020, Astronomy \& Astrophysics, 641, A116, \dodoi{10.1051/0004-6361/202038411}

\bibitem[{Vuitton {et~al.}(2019)Vuitton, Yelle, Klippenstein, H{\"o}rst, \& Lavvas}]{Vuitton.2019}
Vuitton, V., Yelle, R., Klippenstein, S., H{\"o}rst, S., \& Lavvas, P. 2019, Icarus, 324, 120, \dodoi{10.1016/j.icarus.2018.06.013}

\bibitem[{Vuitton {et~al.}(2012)Vuitton, Yelle, Lavvas, \& Klippenstein}]{Vuitton.2012}
Vuitton, V., Yelle, R.~V., Lavvas, P., \& Klippenstein, S.~J. 2012, The Astrophysical Journal, 744, 11, \dodoi{10.1088/0004-637x/744/1/11}

\bibitem[{Waite~Jr. {et~al.}(2007)Waite~Jr., Young, Cravens, Coates, Crary, Magee, \& Westlake}]{Waite.2007}
Waite~Jr., J.~H., Young, D.~T., Cravens, T.~E., {et~al.} 2007, Science, 316, 870–875, \dodoi{10.1126/science.1139727}

\bibitem[{Wright {et~al.}(2024)Wright, Teanby, Irwin, \& Nixon}]{Wright.2024}
Wright, L., Teanby, N.~A., Irwin, P. G.~J., \& Nixon, C.~A. 2024, Experimental Astronomy, 57, 15, \dodoi{10.1007/s10686-024-09934-y}

\bibitem[{Yelle {et~al.}(2008)Yelle, Cui, \& Müller-Wodarg}]{Yelle.2008}
Yelle, R.~V., Cui, J., \& Müller-Wodarg, I. C.~F. 2008, Journal of Geophysical Research: Planets, 113, \dodoi{10.1029/2007je003031}

\end{thebibliography}
\bibliographystyle{aasjournal}



\end{document}